# Study of the rare decays $B^*_{s,d} \to l^+l^-$ in $Z'$ model

**Debika Banerjee** [1], **Priya Maji** [2] **and Sukadev Sahoo** [3]

*Department of Physics, National Institute of Technology, Durgapur-713209, West Bengal, India*

[1] E-mail: rumidebika@gmail.com, [2] E-mail: majipriya@gmail.com,
[3] E-mail: sukadevsahoo@yahoo.com

**Abstract**

The rare decays $B^*_{s,d} \to l^+l^-$ are important to probe the flavour sector of the standard model and to search new physics beyond the SM. Unlike pseudoscalar $B$ meson, the leptonic decays of vector $B^*_{s,d}$ mesons are not chirally supressed which compensates for their short lifetimes, and results significant branching ratios. In this paper, we estimate the branching ratios of $B^*_{s,d} \to l^+l^-$ ($l = e, \mu$) rare decays in $Z'$ model which is an extension of the SM with an extra $U(1)'$ gauge symmetry. We find that the branching ratios are increased from their corresponding standard model values and vary with the mass of $Z'$ boson. Lower is the mass of $Z'$ boson, higher is the branching ratio.

## 1. Introduction

The standard model (SM) of particle physics is the most elegant theory till date with the support of experimental results. Still we find a bunch of open issues associated with the fundamental particles as well as forces which need explanation. Scientists extend their ideas beyond the SM to find the solutions for the unsolved issues such as unification of gravity with other three fundamental forces, presence of tiny neutrino masses, dominance of matter over antimatter in our universe, presence of dark matter and dark energy in our universe etc. Recently, several experimental results from different accelerators have reported several anomalies around $3\sigma$ in $B$ meson sector for observables such as branching ratio of $B_s \to \varphi\mu^+\mu^-$ decay [1], angular observable $P'_5$ in $B \to K^*\mu^+\mu^-$ decay [2], lepton flavour non-universality parameter $R_k$ in $B \to Kl^+l^-$ decay [3]. These anomalies point towards the presence of new physics (NP) effects which need close examination. The rare $B_{s,d} \to l^+l^-$ decays [4-10] play an important role in the SM and the NP sector. In the SM, these decays are loop suppressed and suffer extra helicity suppression. The CMS and LHCb experiments presented their results for the branching ratio of $B_{s,d} \to \mu^+\mu^-$ decays in [11] as: $BR(B_s \to \mu^+\mu^-) = (2.8^{+0.7}_{-0.6}) \times 10^{-9}$ and $BR(B_d \to \mu^+\mu^-) = (3.9^{+1.6}_{-1.4}) \times 10^{-10}$ which are almost in agreement with the SM predictions [12]: $BR(B_s \to \mu^+\mu^-) = (3.65 \pm 0.23) \times 10^{-9}$ and $BR(B_d \to \mu^+\mu^-) = (1.06 \pm 0.09) \times 10^{-10}$. Recently, the ATLAS collaboration has measured the branching ratio $BR(B_s \to \mu^+\mu^-) = (0.9^{+1.1}_{-0.8}) \times 10^{-9}$ and presented an upper limit on $BR(B_d \to e^+e^-) < 4.2 \times 10^{-10}$ at 95% CL [13]. In the SM, the branching ratios for $B_{s,d} \to e^+e^-$ decays are predicted as [12]: $BR(B_s \to e^+e^-) = (8.54 \pm 0.55) \times 10^{-14}$ and $BR(B_d \to e^+e^-) = (2.48 \pm 0.21) \times 10^{-15}$. The CDF collaboration [6] has searched for the $B_{s,d} \to e^+e^-$ decays and obtained the upper limits of branching ratios as: $BR(B_s \to e^+e^-) < 2.8 \times 10^{-7}$ and $BR(B_d \to e^+e^-) < 8.3 \times 10^{-8}$. Due to the presence of large experimental uncertainties we may expect the existence of NP. These $B_{s,d}$ mesons are composite systems; they possess many different energy levels. These energy levels gives rise to a spectrum of excited states [14,15]. The excited mesons $B^*_{s,d}$ are unstable under electromagnetic and strong interactions and possess narrow width with corresponding lifetime of the order of $10^{-17}$ s. The $B^*_{s,d} \to l^+l^-$ decays are sensitive to short-distance structure of $\Delta B=1$ transitions. Thus these decays can be used to test the flavour sector of the SM and search for NP. Khodjamirian et al. [16] proposed a novel method to study flavour-changing neutral currents (FCNCs) in the $B^*_{s,d} \to e^+e^-$ transition and predicted the branching ratio $BR(B^*_{s,d} \to e^+e^-) = 0.98 \times 10^{-11}$. Recently, $B^*_s \to l^+l^-$ decay modes have been studied in the SM in Ref. [17]. They have obtained the branching ratios $BR^{SM}(B^*_{s,d} \to l^+l^-) = (0.7 - 2.2) \times 10^{-11}$ for decay width $\Gamma = 0.10(5)$ keV, irresepective of the lepton flavour. Experimentally, $B^*_{s,d} \to l^+l^-$ decay modes are not observed so far.

In this paper, we study $B^*_{s,d} \to l^+l^-$ ($l = e, \mu$) decays in $Z'$ model. The $Z'$ model arises in the extension of the SM by adding an extra $U(1)'$ gauge symmetry to it. Models beyond the SM predict more than one extra neutral gauge bosons and many new fermions. These new (exotic) fermions can mix with the SM fermions and induce FCNCs [18,19]. Mixing between ordinary (doublet) and exotic singlet left-handed quarks induces FCNC, mediated by the SM $Z$ boson. In these models [20-22], one introduces an additional vector-singlet charge –1/3 quark h, and allows it to mix with the ordinary down-type quarks d, s and b. Since the weak isospin of the exotic quark is different from that of the ordinary quarks, FCNCs involving $Z$ are induced. The $Z$-mediated FCNC couplings $U^Z_{ds}$, $U^Z_{db}$ and $U^Z_{sb}$ are constrained by a variety of processes. The constraints on $U^Z_{db}$ and $U^Z_{sb}$ allow significant contributions to $B_q - \bar{B}_q$ mixing. NP models which contain exotic fermions also predict the existence of additional neutral $Z'$ gauge bosons. The mixing among particles which have different $Z'$ quantum numbers will induce FCNCs due to $Z'$ exchange [23,24]. With FCNCs, the $Z'$ boson contributes at tree level [25,26], and its contribution will interfere with the SM contributions. In this paper, we consider the contribution of $Z'$ boson to $B^*_{s,d} \to l^+l^-$ rare decays and estimate their branching ratios.

This paper is organized in the following manner: In Section 2, we discuss the formalism for $B^*_{s,d} \to l^+l^-$ ($l = e, \mu$) decays in the SM. In Section 3, we discuss $B^*_{s,d} \to l^+l^-$ decays in $Z'$ model. In Section 4, we present our estimated branching ratios numerically as well as graphically for $B^*_{s,d} \to l^+l^-$ deacys in $Z'$ model and compare our results with the SM predictions. We present our conclusion in Section 5.

## 2. $B^*_{s,d} \to l^+l^-$ decays in the SM

The rare leptonic decays $B^*_{s,d} \to l^+l^-$ ($l = e, \mu$) are mediated through $b \to q l^+l^-$ ($q = s, d$) FCNC transitions same as $B_{s,d} \to l^+l^-$ decays. The vector mesons $B^*_{s,d}$ have the same quark content as the $B_{s,d}$ pseudoscalar mesons. The $B_{s,d} \to l^+l^-$ decays experience additional helicity suppression whereas the vector mesons decays $B^*_{s,d} \to l^+l^-$ are free from such helicity suppression. This partially contributes for the shorter lifetime of the $B^*_{s,d}$ and shows the possibility of probing the short-distance structure of the muonic and electronic decays. The SM amplitude for $B^*_{s,d} \to l^+l^-$ decays is given as [17, 27-30]:

$$\text{M} = \frac{G_F}{2\sqrt{2}} V_{tb} V_{tq}^* \frac{\alpha}{\pi} \left[ \begin{array}{c} (m_{B_q^*} f_{B_q^*} C_9 + 2 f^T_{B_q^*} m_b C_7) \bar{l} \not{\varepsilon} l + m_{B_q^*} f_{B_q^*} C_{10} \bar{l} \not{\varepsilon} \gamma_5 l \\ -8\pi^2 \frac{1}{q^2} \sum_{i=1}^{6,8} C_i \langle 0 | T_i^\mu | B_q^*(p,\varepsilon) \rangle \bar{l} \gamma_\mu l \end{array} \right], \quad (1)$$

where $G_F$ is the Fermi coupling constant, $\varepsilon$ is the polarization vector of the $B_q^*$, $f_{B_q^*}$ is decay constant of $B_q^*$ mesons and and $C_i$'s are the Wilson coefficients of the weak Hamiltonian for $\Delta B = 1$ processes [31-33] evaluated at the $b$ quark mass scale [29] in the next-to-next-leading order. The coefficients $C_{9,10}$ are related to the short-distance semileptonic operators and $C_7$ is the coefficient of the electromagnetic penguin operator [34].

The matrix elements of the quark level operators are related to $B_q^*$ meson decay constants as follows:

$$\langle 0 | \bar{q} \gamma^\mu b | B_q^*(p_{B_q^*}, \varepsilon) \rangle = f_{B_q^*} m_{B_q^*} \varepsilon^\mu, \quad (2)$$

$$\langle 0 | \bar{q} \sigma^{\mu\vartheta} b | B_q^*(p_{B_q^*}, \varepsilon) \rangle = -i f^T_{B_q^*} (p^\mu_{B_q^*} \varepsilon^\vartheta - \varepsilon^\mu p^\vartheta_{B_q^*}), \quad (3)$$

$$\langle 0 | \bar{q} \gamma^\mu \gamma_5 b | B_q(p_B) \rangle = -i f_{B_q} p^\mu_{B_q}. \quad (4)$$

The first two matrix elements comes from nonperturbative contributions where $f^T_{B_q^*}$ depends on the renormalization scale. In heavy quark limit $f_{B_q^*}$ are related to $f_{B_q}$ as [35-37]:

$$f_{B_q^*} = f_{B_q}\left(1 - \frac{2\alpha}{3\pi}\right), \quad (5)$$

and $f^T_{B_q^*}$ are related to $f_{B_q}$ as:

$$f^T_{B_q^*} = f_{B_q}\left[1 + \frac{2\alpha}{3\pi}\left(\log\left(\frac{m_b}{\mu}\right) - 1\right)\right]. \quad (6)$$

Considering the renormalization scale at the order of mass of $b$ quark and neglecting the higher order QCD corrections the decay constants $f^T_{B_q^*}$ and $f_{B_q^*}$ reduces to:

$$f_{B_q^*} = f^T_{B_q^*} \simeq f_{B_q}. \quad (7)$$

The corresponding decay width for $B^*_{s,d} \to l^+l^-$ decays is given as:

$$\Gamma = \frac{G_F^2 \alpha^2 |V_{tb} V_{tq}^*|^2}{96\pi^3} m^3_{B_q^*} f^2_{B_q^*} \left[ \left| C_9^{eff}(m^2_{B_q^*}) + 2 \frac{m_b f^T_{B_q^*}}{m_{B_q^*} f_{B_q^*}} C_7^{eff}(m^2_{B_q^*}) \right|^2 + |C_{10}|^2 \right] \quad (8)$$

From equation (8), we observe that the $B^*_{s,d} \to l^+l^-$ decay processes are sensitive to the $C^{eff}_{7,9}$, $C_{10}$ Wilson coefficients, i.e. $O_7$, $O_9$ and $O_{10}$ operators, whereas in the case of $B_{s,d} \to l^+l^-$ decay processes the contributions from $O_7$ and $O_9$ vanish. The values of these branching ratios in the SM are given in Table 1.

**Table 1: The branching ratio of $B^*_{s,d} \to l^+l^-$ ($l = e, \mu$) in the SM [30]**

| Decay Process | Branching Ratio in the SM |
|---|---|
| $B^*_s \to l^+l^-$ | $(1.7 \pm 0.2) \times 10^{-11}$ |
| $B^*_d \to l^+l^-$ | $(1.86 \pm 0.21) \times 10^{-13}$ |

### 3. $B^*_{s,d} \to l^+l^-$ decays in $Z'$ model

In extended quark sector model [20-22, 38], besides the three standard generations of the quarks, there is a $SU(2)_L$ singlet of charge $-1/3$. This model allows for Z-mediated FCNCs. The charged-current interactions are described by

$$L^W_{int} = \frac{g}{\sqrt{2}} \left( W^-_\mu J^{\mu^+} + W^+_\mu J^{\mu^-} \right), \qquad (9)$$

$$J^{\mu^-} = V_{ij} \bar{u}_{iL} \gamma^\mu d_{jL}. \qquad (10)$$

The charged-current mixing matrix V is a 3 × 4 submatrix of K :

$$V_{ij} = K_{ij} \quad \text{for } i = 1,......3, \quad j = 1,......,..4. \qquad (11)$$

Here, V is parameterised by six real angles and three phases, instead of three angles and one phase in the original CKM matrix.

The neutral-current interactions are described by

$$L^Z_{int} = \frac{g}{\cos\theta_W} Z_\mu \left( J^{\mu 3} - \sin^2\theta_W J^\mu_{em} \right), \qquad (12)$$

$$J^{\mu 3} = -\frac{1}{2} U_{pq} \bar{d}_{pL} \gamma^\mu d_{qL} + \frac{1}{2} \delta_{ij} \bar{u}_{iL} \gamma^\mu u_{jL}. \qquad (13)$$

In neutral-current mixing, the matrix for the down sector is U = V$^\dagger$V. Since V is not unitary, $U \neq 1$, the nondiagonal elements do not vanish:

$$U_{pq} = - K^*_{4p} K_{4q} \quad \text{for } p \neq q. \qquad (14)$$

The various $U_{pq}$ are non-vanishing, which allow for flavour-changing neutral currents that would be a signal for new physics.

Now consider the $B^*_{s,d} \to l^+l^-$ decays in the presence of Z-mediated FCNC. The $B^*_{s,d} \to l^+l^-$ decays are mediated through the same FCNC transitions as $B_{s,d} \to l^+l^-$ decays.

Therefore, the expressions for effective Hamiltonian of $B_{s,d}^* \to l^+l^-$ decays can be written analogous to $B_{s,d} \to l^+l^-$ decays [7]. Considering the contribution of Z-mediated FCNC to $B_q^* \to l^+l^-$ ($q = s, d$ and $l = e, \mu$), one can write the effective Hamiltonian as:

$$H_{eff}(Z) = \frac{G_F}{\sqrt{2}} U_{qb} \left[\bar{q}\gamma^\mu(1-\gamma_5)b\right]\left[\bar{\ell}\left(C_V^\ell \gamma_\mu - C_A^\ell \gamma_\mu \gamma_5\right)\ell\right], \tag{15}$$

where $C_V^\ell$ and $C_A^\ell$ are the vector and axial vector $Z\ell^+\ell^-$ couplings and are given as

$$C_V^\ell = -\frac{1}{2} + 2\sin^2\theta_W, \quad C_A^\ell = -\frac{1}{2}. \tag{16}$$

We can apply the same concept for a $Z'$ boson i.e., mixing among particles which have different $Z'$ quantum numbers will induce FCNCs due to $Z'$ exchange [23-24, 39-42] and these effects can be as large as Z-mediated FCNCs. However, the $Z'$-mediated coupling $U_{pq}^{Z'}$ can be generated via mixing of particles with same weak isospin and are not suppressed by the mass of heavy fermion. Even though $Z'$-mediated interactions are suppressed relative to Z, these are compensated by the factor $U_{pq}^{Z'}/U_{pq}^Z \sim (M_2/M_1)$. Hence the effect of $Z'$-mediated FCNCs are comparable to that of Z-mediated FCNCs. If we assume $\left|U_{qb}^{Z'}\right| \sim \left|V_{tb}V_{tq}^*\right|$, then it is possible to write $U_{qb}$ instead of $U_{qb}^{Z'}$, which gives significant contributions to the $B_{s,d}^* \to l^+l^-$ decay. The new contributions from $Z'$ boson have similar effect as from the Z boson. Therefore, we write the general effective Hamiltonian that contributes to $B_q^* \to l^+l^-$, in the light of equation (15) as:

$$H_{eff}(Z') = \frac{G_F}{\sqrt{2}} U_{qb} \left[\bar{q}\gamma^\mu(1-\gamma_5)b\right]\left[\bar{\ell}\left(C_V^\ell \gamma_\mu - C_A^\ell \gamma_\mu \gamma_5\right)\ell\right]\left(\frac{g'}{g}\frac{M_Z}{M_{Z'}}\right)^2, \tag{17}$$

where $g = e/(\sin\theta_W \cos\theta_W)$ and $g'$ is the gauge coupling associated with the $U(1)'$ group. The net effective Hamiltonian can be written, from equation (15) and (17), as $H_{eff} = H_{eff}(Z) + H_{eff}(Z')$ and

$$H_{eff} = \frac{G_F}{\sqrt{2}} U_{qb} \left[\bar{q}\gamma^\mu(1-\gamma_5)b\right]\left[\bar{\ell}\left(C_V^\ell \gamma_\mu - C_A^\ell \gamma_\mu \gamma_5\right)\ell\right]\left[1+\left(\frac{g'}{g}\frac{M_Z}{M_{Z'}}\right)^2\right]. \tag{18}$$

### 4. Results and Discussions

In this section, we calculate the branching ratios for $B_{s,d}^* \to l^+l^-$ ($l = e, \mu$) rare decays using recent data from PDG [43]. The value of $g'/g$ is undetermined [44]. However, generically, one expects that $g'/g \approx 1$ if both $U(1)$ groups have the same origin from some grand unified theory. We take $g'/g \approx 1$ in our calculations. The $Z'$ boson has not yet been discovered, so its exact mass is unknown. The $Z'$ mass is constrained by direct searches from different accelerators [45-46], which give a model-dependent lower bound around 500 GeV. Sahoo *et* *al.* [47] estimated $Z'$ boson mass from $B_q^0 - \overline{B_q^0}$ mixing which lies in the range of 1352–1665 GeV. Oda et al. [48] have predicted an upper bound on $Z'$ boson mas, $M_{Z'} \leq 6$ TeV in classically conformal $U(1)'$ extended standard model. The ATLAS collaboration [49] sets the lower mass limits for the sequential standard model (SSM) $Z'_{SSM}$ as 1.90 TeV and ranges from 1.82 – 2.17 TeV are excluded for a $Z'_{SFM}$ strong flavor model. Recently, the CMS collaboration [50] has searched leptophobic $Z'$ bosons decaying into four-lepton final states in proton-proton collisions $\sqrt{s} = 8$ TeV and obtained the lower limit on the $Z'$ boson mass as 2.5 TeV. In this work, we have used the lower limit of $M_{Z'} = 500$ GeV and upper limit $M_{Z'} = 6$ TeV for our calculations. We estimate the value of branching ratios corresponding to different mass of $Z'$. The calculated values are encapsulated in Table 2. The dependence of branching ratio for $B_{s,d}^* \to l^+l^-$ ($l = e, \mu$) decays on $M_{Z'}$ is shown in Fig. 1 and Fig. 2 respectively. We find that the branching ratios are increased from their corresponding SM values [30]. Furthermore, the figures show that depending on the precise value of $M_{Z'}$, $Z'$-mediated FCNCs give sizable contributions to $B_{s,d}^* \to l^+l^-$ decays. Lower is the mass of $Z'$ boson, more is the contribution towards the branching ratio. Our predicted branching ratios for $B_{s,d}^* \to \mu^+\mu^-$ decays are approximately two to three orders less than the branching ratios of $B_{s,d} \to \mu^+\mu^-$ processes, whereas the branching ratios for $B_{s,d}^* \to e^+e^-$ decays are approximately two to three orders more than the branching ratios of $B_{s,d} \to e^+e^-$ processes in the SM. We expect that the sensitivity of $Z'$ model is different for different lepton flavours in excited B meson decays. So these rare decays may be used to test the lepton flavour universality. The experimental measurements for $B_{s,d}^* \to l^+l^-$ decay processes will clear the conjecture between SM predictions and new physics scenarios.

**Table 2: Numerical estimation of the *BR* of $B_{s,d}^* \to l^+l^-$ ($l = e, \mu$) in $Z'$ model.**

| Decay Process | $M_{Z'}$ in TeV | $BR|_{SM+Z'}$ |
|---|---|---|
| $B_s^* \to l^+l^-$ | 0.5 | $(1.601 - 2.208) \times 10^{-11}$ |
| | 1 | $(1.525 - 1.931) \times 10^{-11}$ |
| | 4 | $(1.501 - 1.901) \times 10^{-11}$ |
| | 6 | $(1.500 - 1.900) \times 10^{-11}$ |
| $B_d^* \to l^+l^-$ | 0.5 | $(1.761 - 2.209) \times 10^{-13}$ |
| | 1 | $(1.677 - 2.104) \times 10^{-13}$ |
| | 4 | $(1.651 - 2.072) \times 10^{-13}$ |
| | 6 | $(1.650 - 2.071) \times 10^{-13}$ |

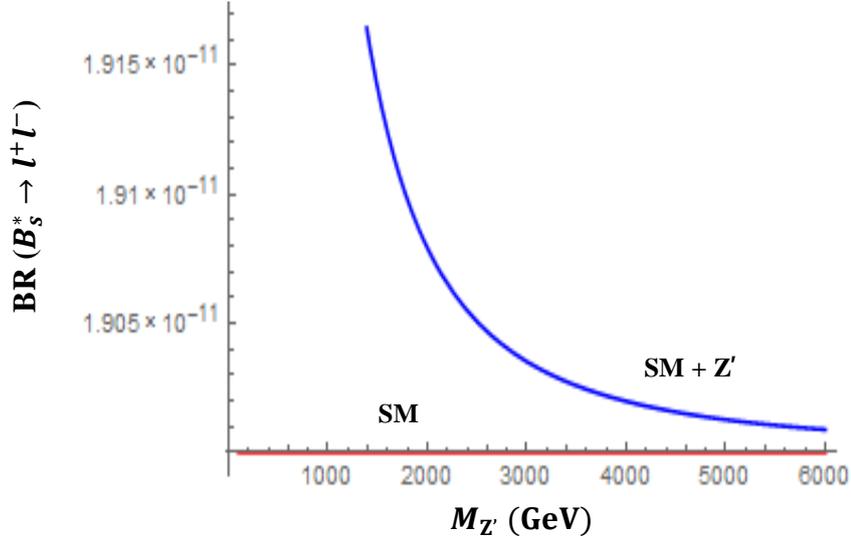

Fig.1: The variation of branching ratio $B_s^* \to l^+l^-$ with the mass of $Z'$ boson.

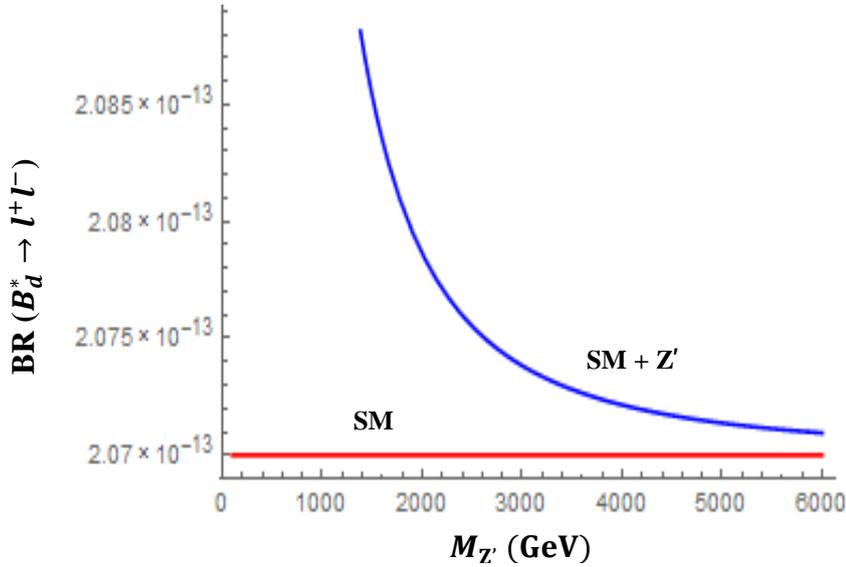

Fig. 2: The variation of branching ratio $B_d^* \to l^+l^-$ with the mass of $Z'$ boson.

## 5. Conclusion

The $B_{s,d}^* \to l^+l^-$ ($l = e, \mu$) rare decays play an important role in the standard model and their study is relevant to indirect searches for physics beyond the standard model. These decays are theoretically very clean because their amplitudes only depend on $B_{s,d}^*$ meson decay constants which can be determined using non-perturbative methods such as QCD sum rules, lattice gauge theory and so on. These decays can provide an excellent environment for giving complimentary information on the semileptonic $b \to (s,d)l^+l^-$ operators. Recently, $B_{s,d}^* \to l^+l^-$ decay rates and branching ratios have been predicted in the SM [17]. However, $B_{s,d}^* \to l^+l^-$ decays are not observed experimentally so far. In this paper, we study the effect of $Z'$-mediated FCNCs on these decays. We have found that although there is no noticeable difference in the branching ratios for $B_{s,d}^* \to l^+l^-$ decays between the SM values and the values in $Z'$ model, the branching ratio values increase from their SM values. Furthermore, the branching ratio of $B_{s,d}^* \to l^+l^-$ decays varies with the mass of $Z'$ boson. Lower is the mass of $Z'$ boson, higher is the branching ratio. We expect these rare decays provide very useful tool for affording new tests of lepton universality and to explore new physics beyond the SM. If these branching ratios are measured in future, their precise measurement would limit the allowed range of the mass of $Z'$ boson. We are hoping for a precise measurement of these branching ratios at the LHC or any of the future colliders.

**Acknowledgments**

We thank the reviewer for suggesting valuable improvements of our manuscript. D. Banerjee and P. Maji acknowledge the Department of Science and Technology, Govt. of India for providing INSPIRE Fellowship through IF140258 and IF160115 respectively for their research work. S. Sahoo would like to thank Science and Engineering Research Board (SERB), Department of Science and Technology, Govt. of India for financial support through grant no. EMR/2015/000817. S. Sahoo also acknowledges the financial support of NIT Durgapur through "Research Initiation Grant" office order No. NITD/Regis/OR/25 dated 31st March, 2014.